\documentclass{sig-alternate}

\usepackage{array}
\usepackage{multirow}
\usepackage{listings} 
\usepackage{color}
\usepackage{epstopdf}
\definecolor{mygray}{gray}{0.98}

\begin{document}
%
\conferenceinfo{EWILI 2013, 26-27 August 2013, Toulouse, France}{}
\CopyrightYear{2013} 

\title{DyPS: Dynamic Processor Switching for Energy-Aware Video Decoding on Multi-core SoCs}
%
%
%
%
%

\numberofauthors{5} 
%
\author{
%
%
\alignauthor
Yahia Benmoussa\\
       \affaddr{Univ. M'hamed Bougara, LMSS, Boumerdes, Algeria\\}
	\affaddr{Univ. Bretagne Occidentale, UMR6285, Lab-STICC, F29200 Brest, France }
       \email{yahia.benmoussa@univ-brest.fr}
\alignauthor
Jalil Boukhobza\\
	\affaddr{Univ. Bretagne Occidentale, UMR6285, Lab-STICC, F29200 Brest, France}
       \email{jalil.boukhobza@univ-brest.fr}
\alignauthor
Eric Senn\\
       \affaddr{Univ. Bretagne Sud, UMR6285, Lab-STICC, F56100 Lorient, France}\\
       \email{eric.senn@univ-ubs.fr}
\and  
\alignauthor
Djamel Benazzouz\\
       \affaddr{Univ. M'hamed Bougara, LMSS, Boumerdes, Algeria}\\
       \email{dbenazzouz@yahoo.fr}
\alignauthor Yassine Hadjadj-Aoul\\
      \affaddr{IRISA, Universit\'e de Rennes1}\\
       \email{yassine.hadjadj-aoul@irisa.fr}
       }
  
\maketitle
\begin{abstract}

In addition to General Purpose Processors (GPP), Multi-core SoCs equipping modern mobile devices contain specialized Digital Signal Processor designed with the aim to provide better performance and low energy consumption properties. However, the experimental measurements we have achieved revealed that system overhead, in case of DSP video decoding,  causes drastic  performances drop  and energy efficiency as compared to the GPP decoding. This paper describes DyPS, a new approach for energy-aware processor switching (GPP or DSP) according to the video quality. We show the pertinence of our solution in the context of adaptive video decoding  and describe an implementation on an embedded Linux operating system with the help of the GStreamer framework. A simple case study showed that DyPS achieves 30\% energy saving while sustaining the decoding performance. 

\end{abstract}




\section{Introduction}
Nowadays, mobile devices such as smart-phones and tablets include more and more powerful hardware. For example, a hardware configuration including  a  processor clocked at more than 1 GHz frequency becomes common. However the use of high frequencies requires higher voltage levels and  leads to an increase in energy consumption due to the quadratic relation between the dynamic power and the supplied voltage in CMOS circuits \cite{375385}. In a context where Lithium battery technologies are not evolving fast enough, the autonomy duration  of those devices becomes a very critical issue \cite{broussely_li-ion_2004} especially when using processor intensive applications such as video playback.  In \cite{carroll2010analysis}, it is shown that video playback is the most important energy intensive mobile application. This is due to the important use of the processing resources responsible of more than 60\% of the consumed energy.

To overcome this issue,  Digital Signal Processors (DSP) are a solution used  to provide better performance-energy properties. Indeed, the use of parallelism in data processing increases the performance without requiring higher voltages and frequencies \cite{1317052}. This makes them an energy-efficient choice in energy constrained devices \cite{1012351} such as smart-phones and tablets where they are  integrated in multi-core SoCs in addition to GPP \cite{vanBerkel:2009:MMP:1874620.1874924}. 

When decoding a video stream,  the use of the full processing capabilities of the  hardware is not always necessary. For example, due to bandwidth limitation, the video may be coded in a low quality which leads to less decoding processing requirements \cite{1218201}. In this case, in order to save energy, one might use dynamic voltage and frequency scaling feature provided by some low-power processors. This mechanism is used to scale down the voltage and the frequency in case of low  processing workloads  \cite{Pouwelse:2001:DVS:381677.381701}.

In addition to the above-stated energy considerations, the operation system overhead is an important parameter to  consider especially in case of DSP video coding. In fact, the inter-processor communication generates a system overhead resulting from cache memory coherency maintenance, parameters passing, and I/O latency. This overhead is not negligible in case of decoding a low  quality video requiring less processing power. In such case, as confirmed by  experimental performance and energy consumption measurements \cite{ARM-vs-DSP-MASCOTS}, it was shown that a GPP video decoding can be the best choice in many cases as compared to the DSP decoding. 

Accordingly, we propose in this paper an implementation of an energy-aware dynamic processing resources selection technique. This new approach allows a transparent processor switching (DSP/GPP) on a multi-core SoCs including a GPP and a DSP in a context of adaptive  decoding of different video qualities.

The remainder of this paper is organized as follows : In section \ref{backgroud}, the problem statement and context are given.  In sections \ref{solution},  the proposed solution is described.  The implementation details,  experimental evaluation and results are discussed in section \ref{implementation} and \ref{results} respectively. Related works on energy consideration of video decoding on Multi-core SoCs are discussed in section \ref{related}. Finally, conclusions and some future work perspectives are given in Section \ref{conclusion}.

\section{Problem statement and context}
\label{backgroud}

\subsection{Problem Statement}

\begin{figure}[!h]
\centering
\includegraphics[width=3.3in]{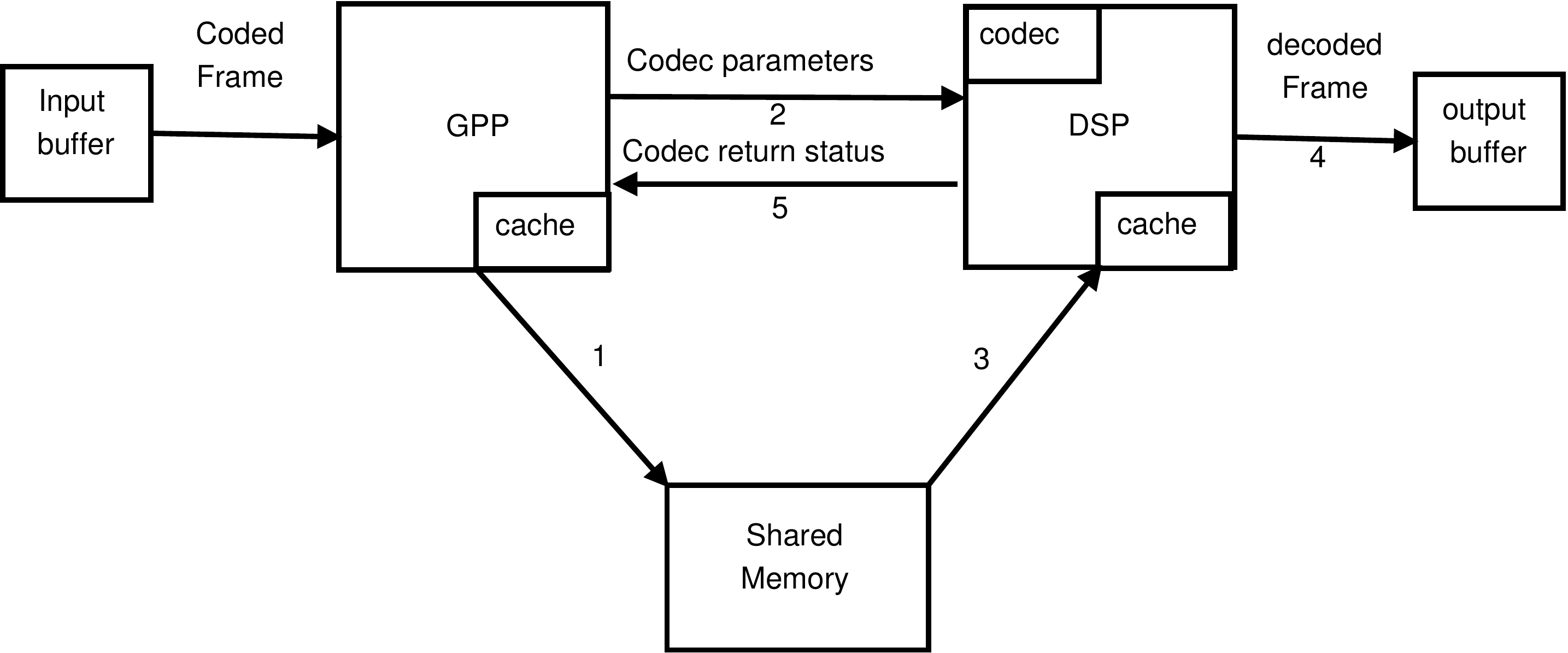}
 \caption{DSP video decoding}
 \label{DSP-video-decoding}
\end{figure}

When decoding a video stream using a DSP, the inter-processor communication may generate a system overhead, and thus  additional energy consumption \cite{DSP-overhead-article, CodecEngineOverhead}. As an illustration, Fig. \ref{DSP-video-decoding} describes the steps of a typical DSP video decoding process controlled by a GPP. The video frames are supposed in an input buffer in the memory. 
\label{dsp-overhead}

\begin{enumerate}
 \item The GPP writes-back the frame located in its cache (as the frame may be located in the cache) to a shared memory so that the DSP can access it.
\item The GPP sends the parameters to the DSP codec via a GPP/DSP hardware bus.
\item The DSP invalidates the entries in its cache corresponding to a frame buffer in the shared memory.
\item The DSP decodes and transfers the frame to the output buffer.
\item The DSP sends the return status to the GPP.  
\end{enumerate}

In fact, both  DSP and GPP have their proper cache memory and communicate using a shared memory. This imposes to manage cache coherency each time the DSP shares a data with the GPP. In addition, from the operating system level, The GPP/DSP communication is managed by a driver. A frame (a compressed picture) decoding is considered, from the GPP point of view, as an I/O operation generating a system latency caused by entering the \textit{idle} state and handling of hardware interrupt. In addition, the GPP/DSP data transfers are generally achieved using Direct Memory Access (DMA) which  offloads the processor from memory data transfers tasks but induces additional I/O and interrupt latency. 

\begin{figure}[!t]	
\centering
\includegraphics[width=3in]{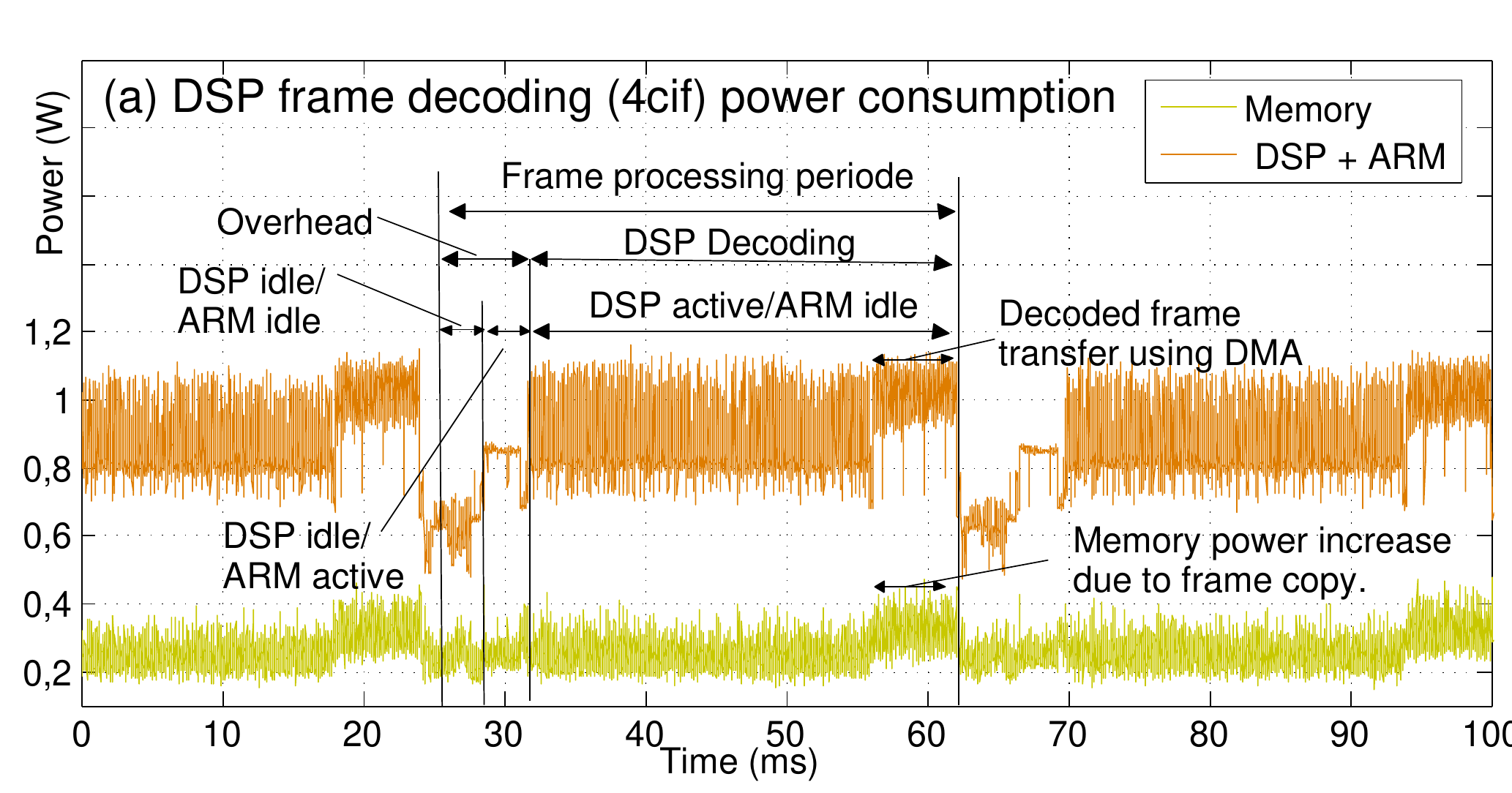}
\includegraphics[width=3in]{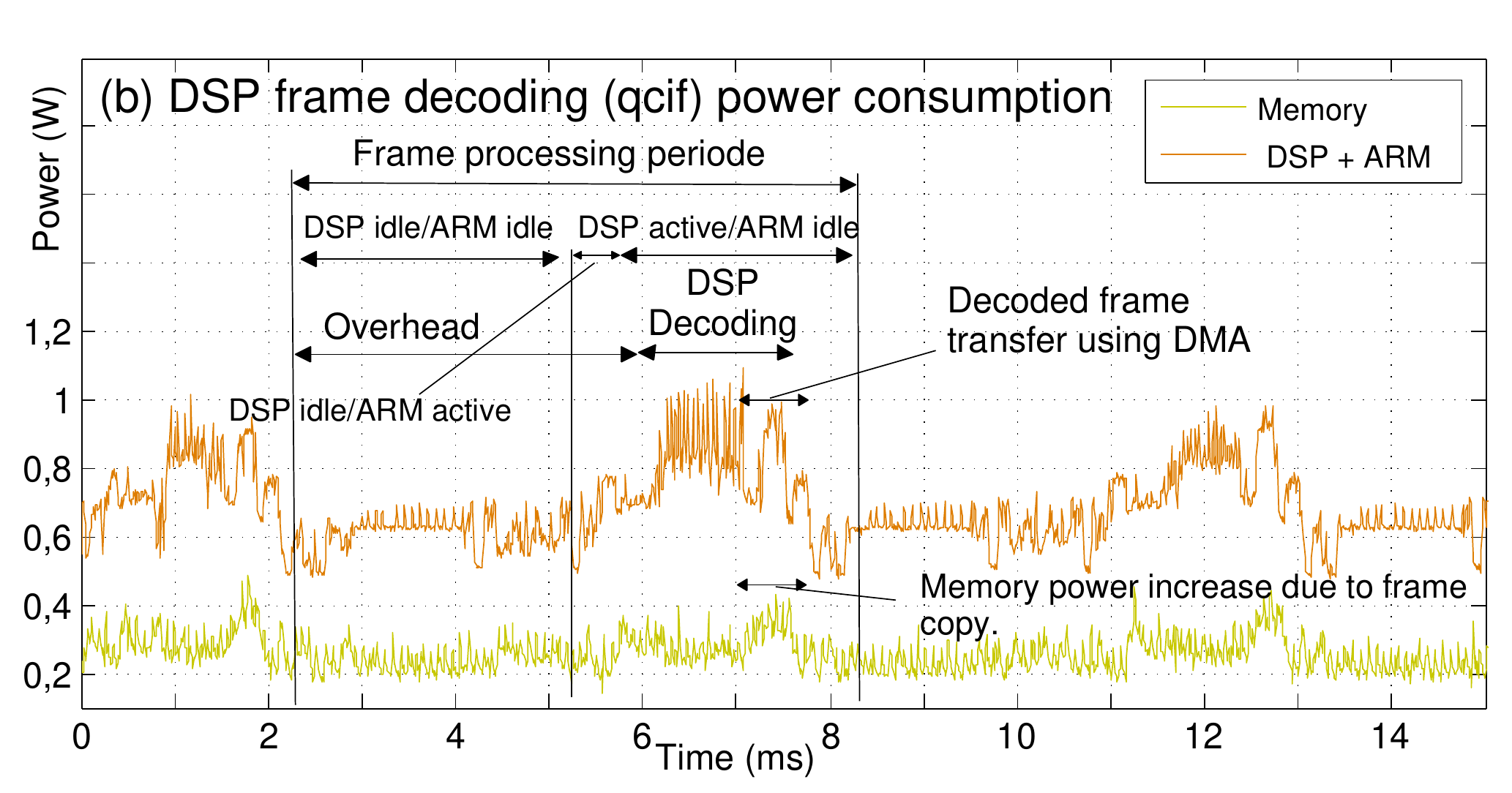}
\caption{ARM and DSP frame decoding}
\label{frame-dec}
\end{figure}

\begin{figure*}[!t]
\centering
\includegraphics[width=1.6in]{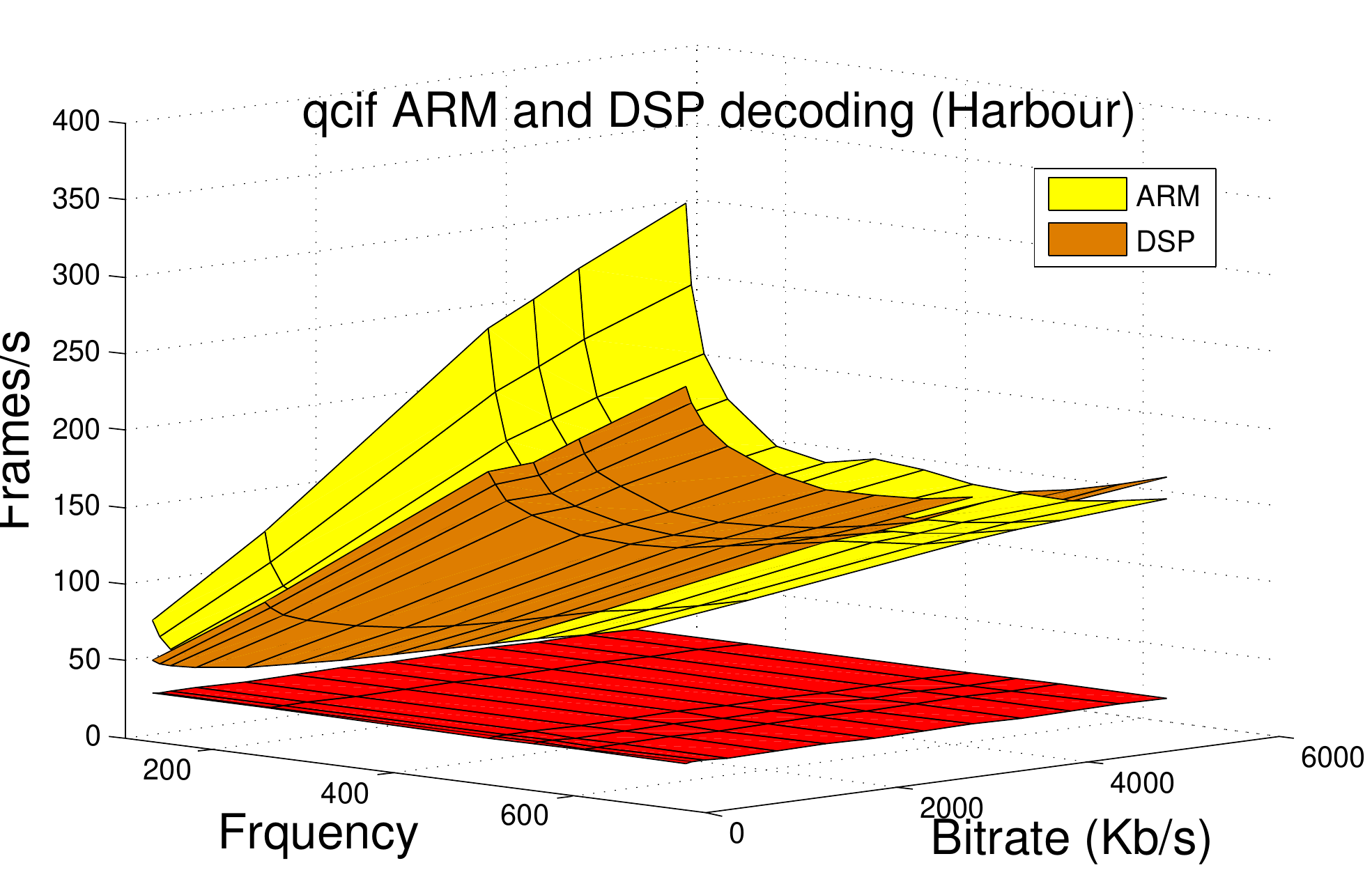}
\includegraphics[width=1.6in]{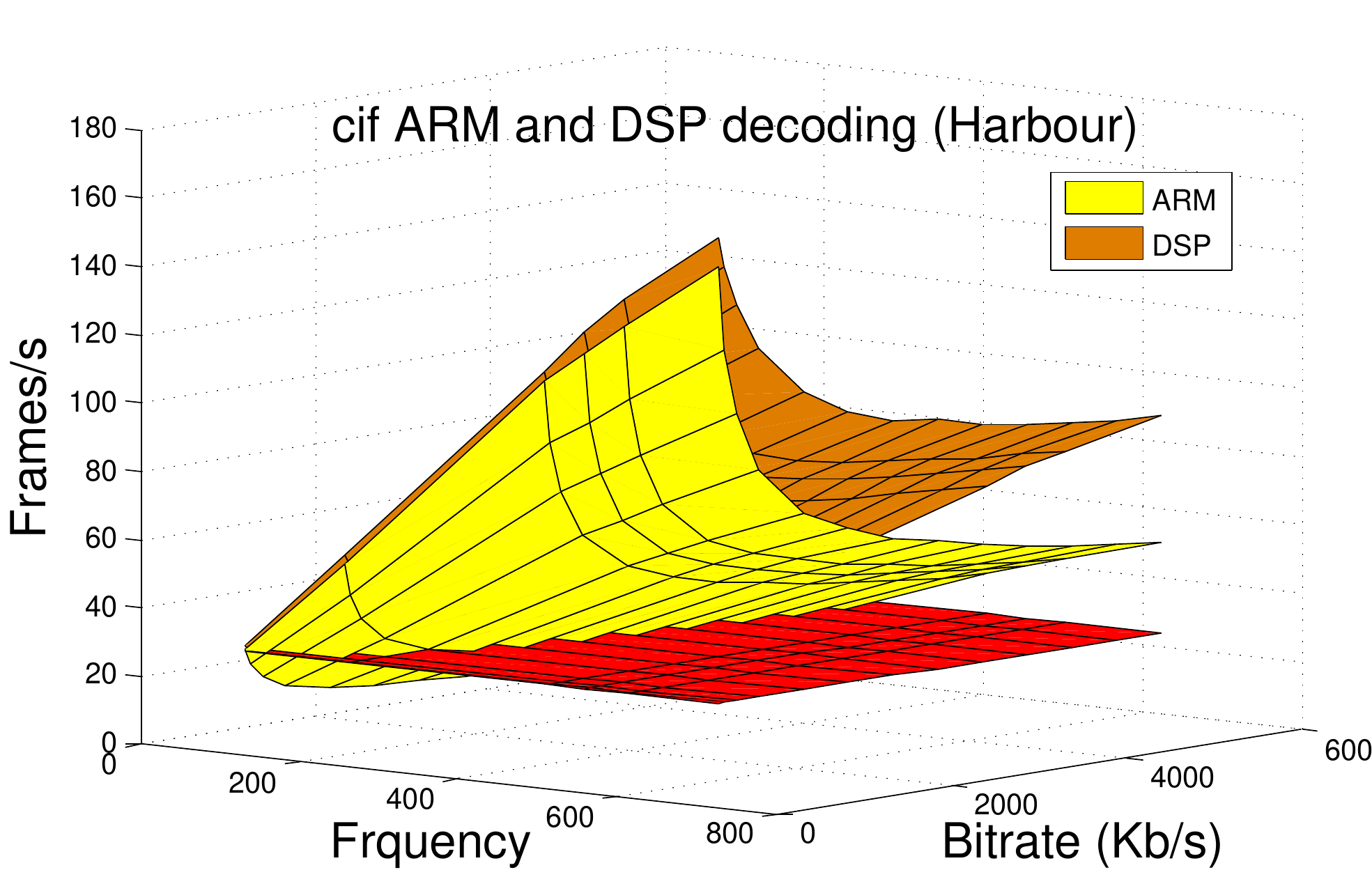}
\includegraphics[width=1.6in]{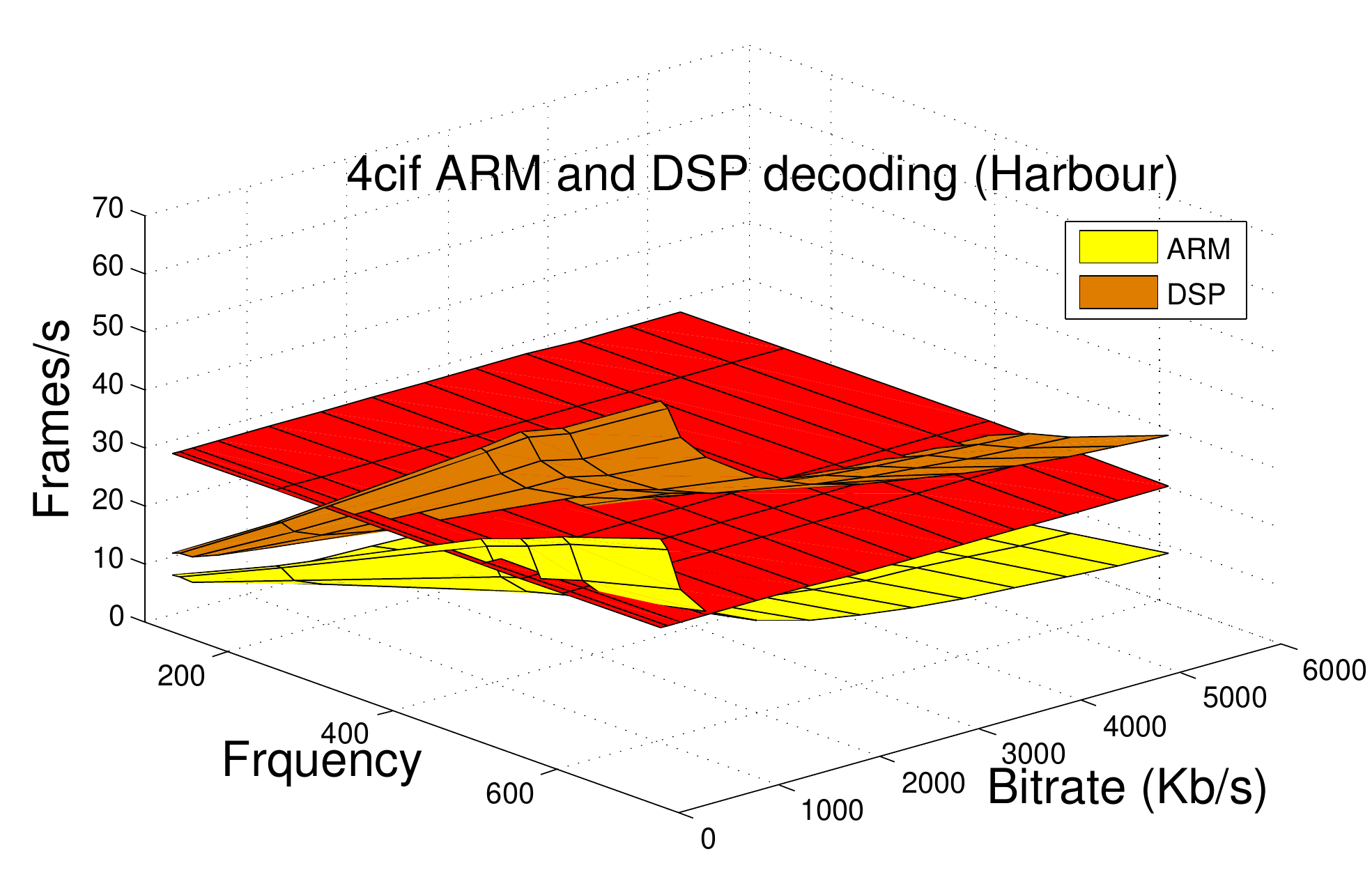}
 \caption{ARM \textit{vs} DSP video decoding performance}
\label{3d-fps}
\end{figure*}
   
\begin{figure*}[!t]
\centering
\includegraphics[width=1.6in]{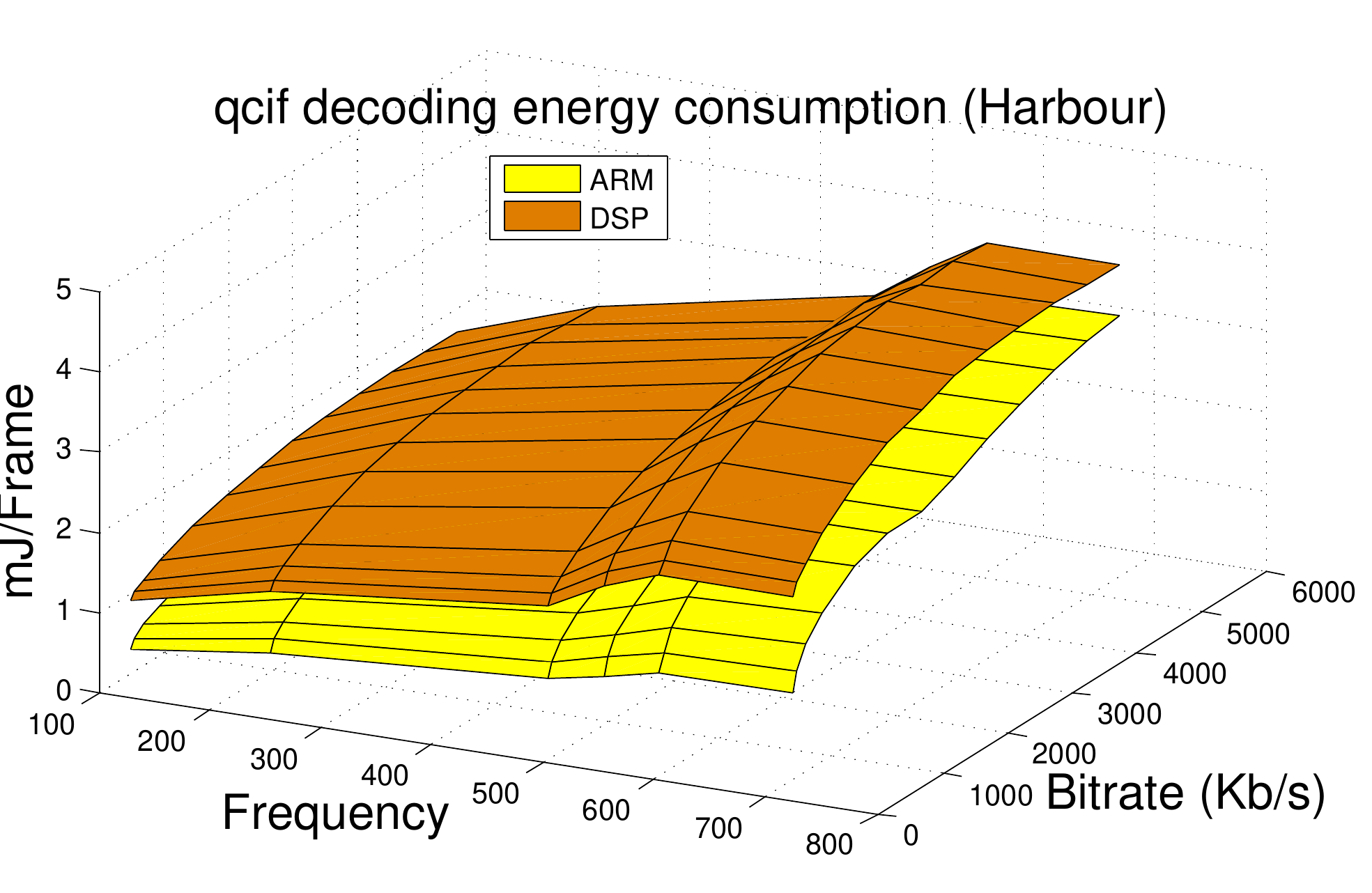}
\includegraphics[width=1.6in]{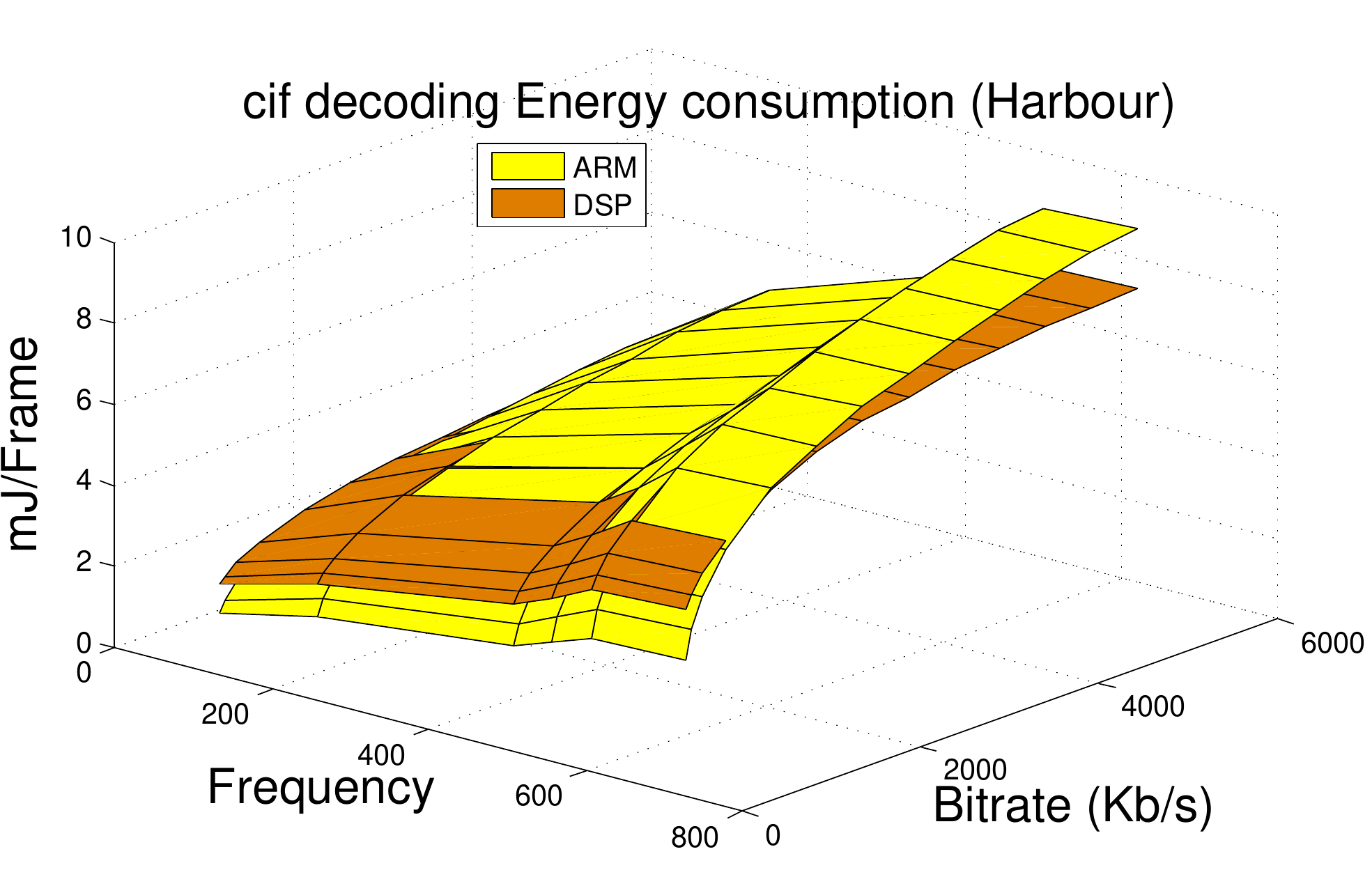}
\includegraphics[width=1.6in]{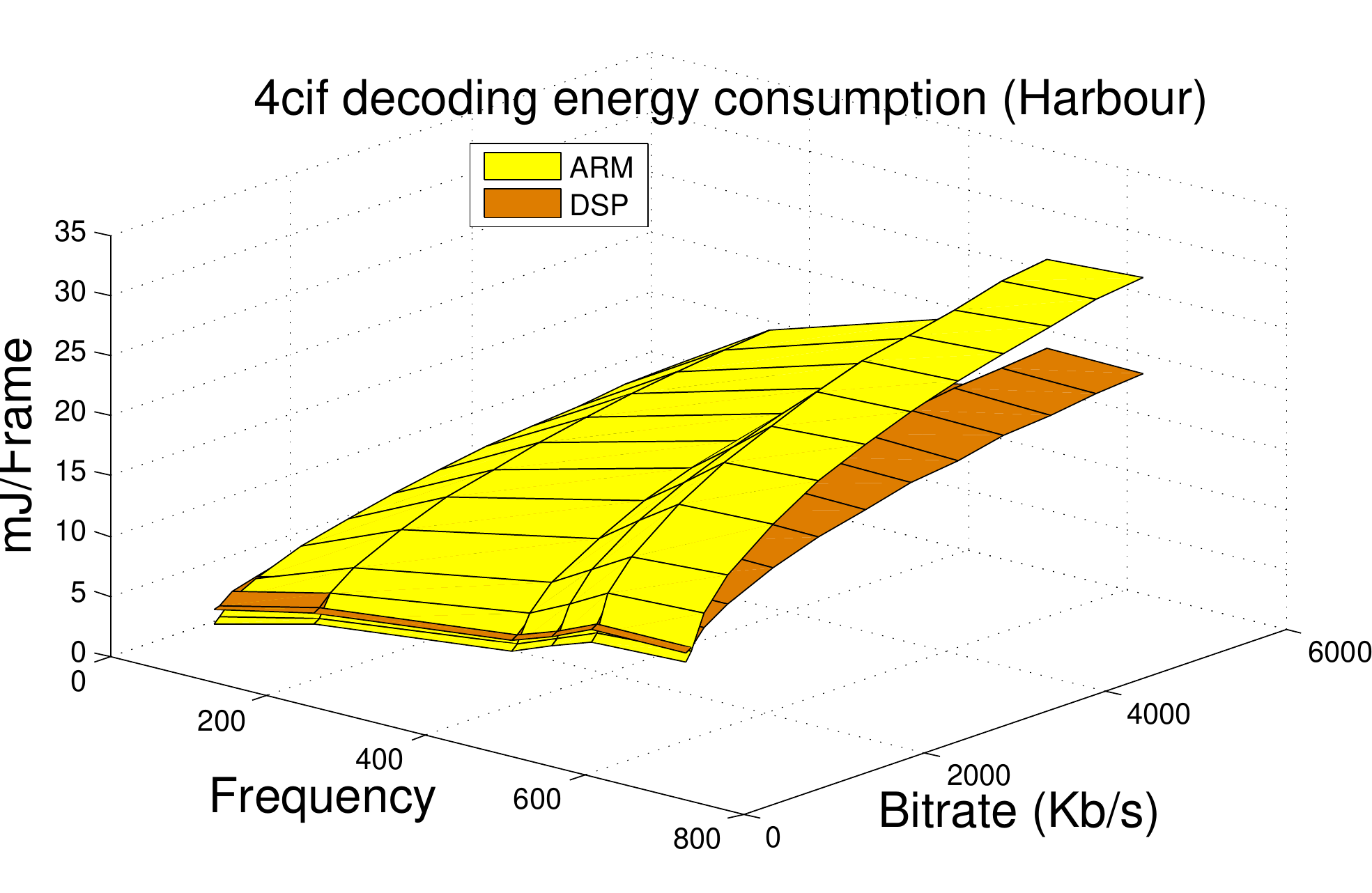}
\caption{ARM vs DSP decoding energy consumption of H264/AVC video}
\label{3d-energy}
\end{figure*}

A frame based DSP decoding analysis showed that the system overhead cannot be neglected  in case of low video  quality DSP decoding \cite{ARM-vs-DSP-MASCOTS}. For example, Figure \ref{frame-dec}  shows the measured  power consumption  during the execution of the above steps on an OMAP3530 SoC containing a Cortex A8 ARM processor and a TMS320C64X DSP. Two video qualities are used : \textit{4cif} (704x576) and {qcif} (176x144) resolution with 4 Mb/s and 256 Kb/s bit-rate respectively. The ARM processor and the DSP are clocked at 720 MHz and 520 MHz frequency respectively.  

In Figure \ref{frame-dec}, the DSP frame decoding phase is represented by the strip varying between 0.7 W and 1.1 W corresponding to [32 ms, 62ms] and [6.2 ms, 7.5ms] intervals (\textit{4cif} and \textit{qcif} respectively). This phase  is terminated by a burst of DMA transfers of the decoded frame macro-blocks from the DSP cache to the shared memory. This phase corresponds to the intervals [56 ms, 62 ms] and [7.2 ms, 7.5 ms] and is illustrated by an increase in memory power consumption. When the DSP terminates the frame decoding, it returns to the GPP (ARM Cortex A8) the execution status and enters the \textit{idle} state. This event occurs, for example in Fig. \ref{frame-dec}-a (\textit{4cif}) at 25 ms. The ARM wakeup latency is represented  by the power level 0.66 W. The ARM wakeup event is represented by the power transition from 0.64 W to 0.85 W level.

A deeper performance and energy measurement showed that the processing which is not related to frame decoding (system overhead) represents 50\% of the total processing time and 30\% of the consumed energy in case of \textit{qcif} resolution. This is not negligible and may have an impact on the overall performance and energy properties of video decoding we discuss hereafter.

\subsubsection{Performances Impact}

Figure \ref{3d-fps} shows, a comparison between the measured performance (Decoded Frames/s) of GPP and DSP video decoding (Harbor sequence/30 Hz) according to the video bit-rate, resolution and clock frequency. The flat surface is the reference  decoding speed  corresponding to the  display rate (30 Hz). It appears clearly that the DSP performances drops as compared to ARM decoding in case of \textit{qcif} resolution.

\subsubsection{Energy consumption Impact}

Figure \ref{3d-energy} shows a comparison between the energy consumption (mJ/Frame) of ARM and DSP video decoding for the same sequence according to the video bit-rate, resolution and processor frequency. One can observe that in case of \textit{qcif} and low bit-rate \textit{cif} resolution (352x288), the ARM video decoding is more energy-efficient than the DSP. 

More information on performance and energy behavior of DSP decoding in term of video quality can be found in  \cite{ARM-vs-DSP-MASCOTS}. In the next section, we discuss the opportunity to exploit these results in a adaptive video decoding context.  

\subsection{Context}

\begin{figure}[!h]
\centering
\includegraphics[width=2.8in]{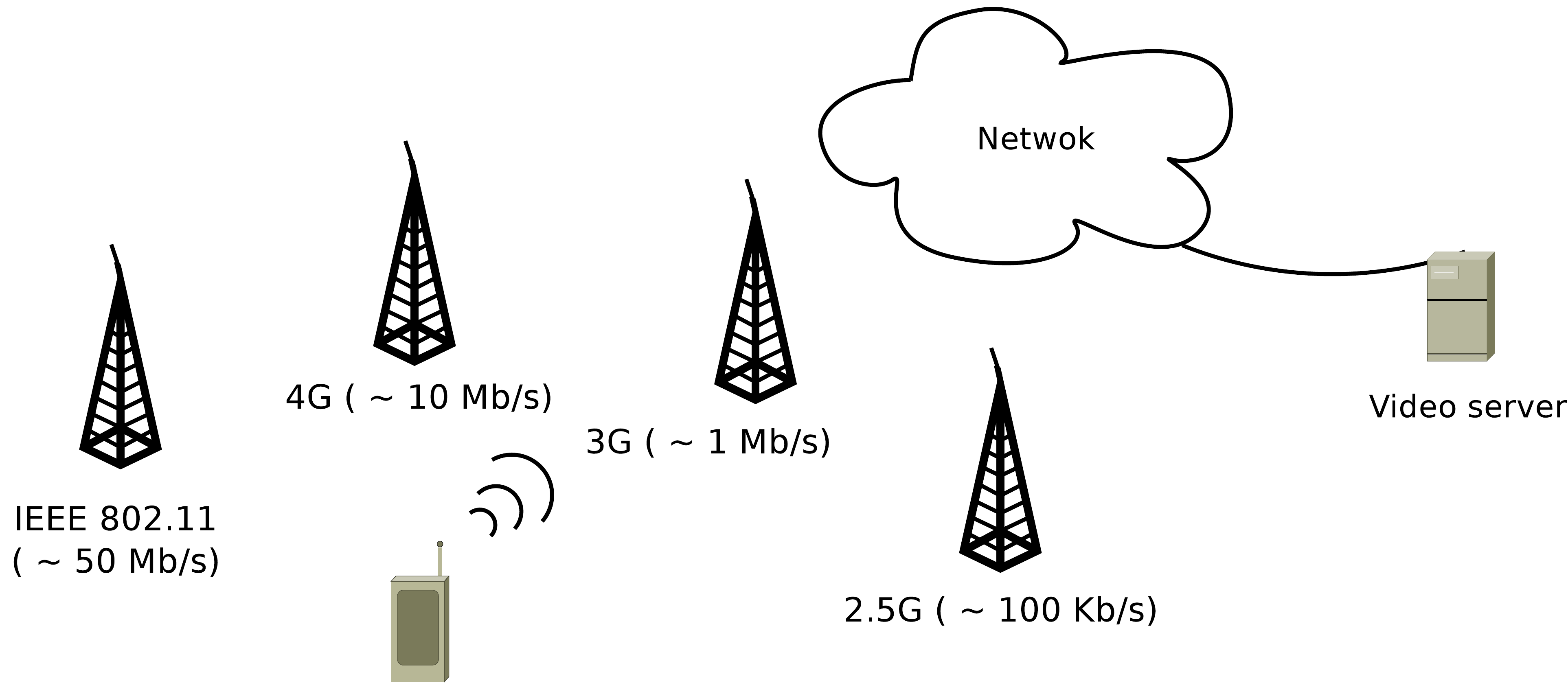}
\caption{Video delivery via heterogeneous networks}
\label{network}
\end{figure}

In mobile devices, a video content may be accessed using heterogeneous networks. Figure \ref{network} illustrates an examples of some network technologies and their bandwidth capabilities which range from tens of Kbits to tens of Mbits per seconds. Consequently, the video  quality should vary when the mobile device roams from a network to another. In addition, the bandwidth may fluctuate within the same network due to network  congestion. In this context, novel streaming and video coding standards  \cite{Stockhammer:2011:DAS:1943552.1943572,schwarz_overview_2007} are designed to adapt dynamically the video content quality to fit with the available networks bandwidth. Based on the aforementioned observations, one can leverage such dynamic video adaptation  by  selecting  the best suitable processing resource (GPP or DSP) available on multi-core SoCs for a better energy-efficiency.

\section{Proposed solution}
\label{solution}

We designed here a solution  that consists in selecting the best processing resource in the context of adaptive video decoding on multi-core SoC containing a GPP and DSP. 

\subsection{Scope}
\label{scope}
Two video quality adaptation approaches exist: The scalable video coding \cite{schwarz_overview_2007}  and adaptive streaming \cite{Stockhammer:2011:DAS:1943552.1943572}. In the first one, a video content is coded in a self-contained  file containing multiple layers : a base video quality layer and multiple enhancement layers. The enhancement layer consists of incremental data which allow to obtain a higher video quality starting from the base layer. In this case, there is a dependency between the different video qualities. 

\begin{figure}[!h]
\centering
\includegraphics[width=2.8in]{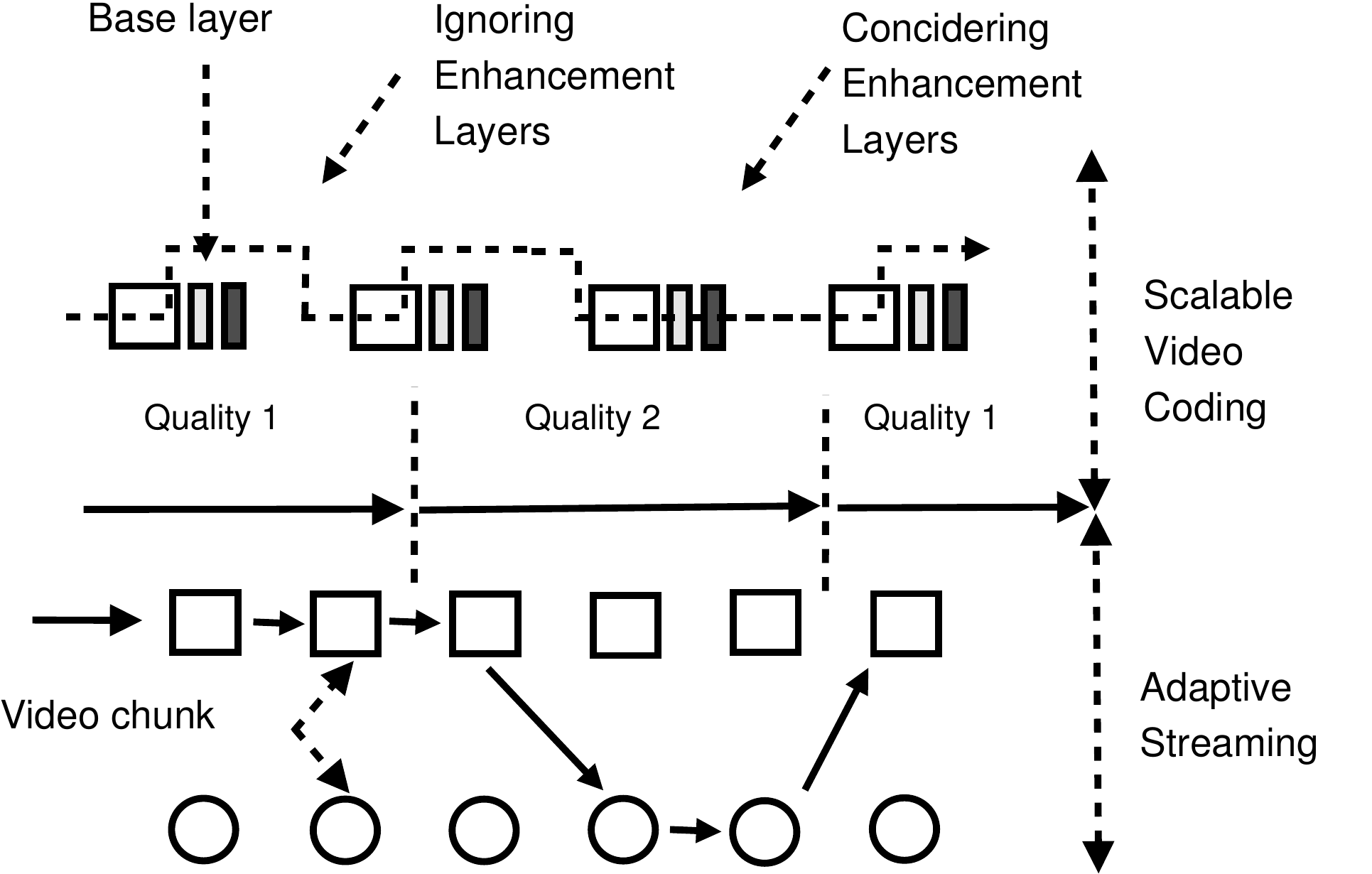}
 \caption{Scalable video coding and adaptive streaming}
	 \label{adaptive-vs-scalable}
\end{figure}

On the other hand, in adaptive streaming, a single video content is coded in independent streams having different qualities. each stream is divided into smaller units (video  sequences of few seconds) called chunks. The quality adaptation is achieved at a chunk granularity. Each chunk is decoded independently from the other chunks. Figure \ref{adaptive-vs-scalable} illustrates these two approaches.

\subsection{Adaptation logic}
\label{adaptation-logic}

\begin{figure}[!h]
\centering
\includegraphics[width=2.3in]{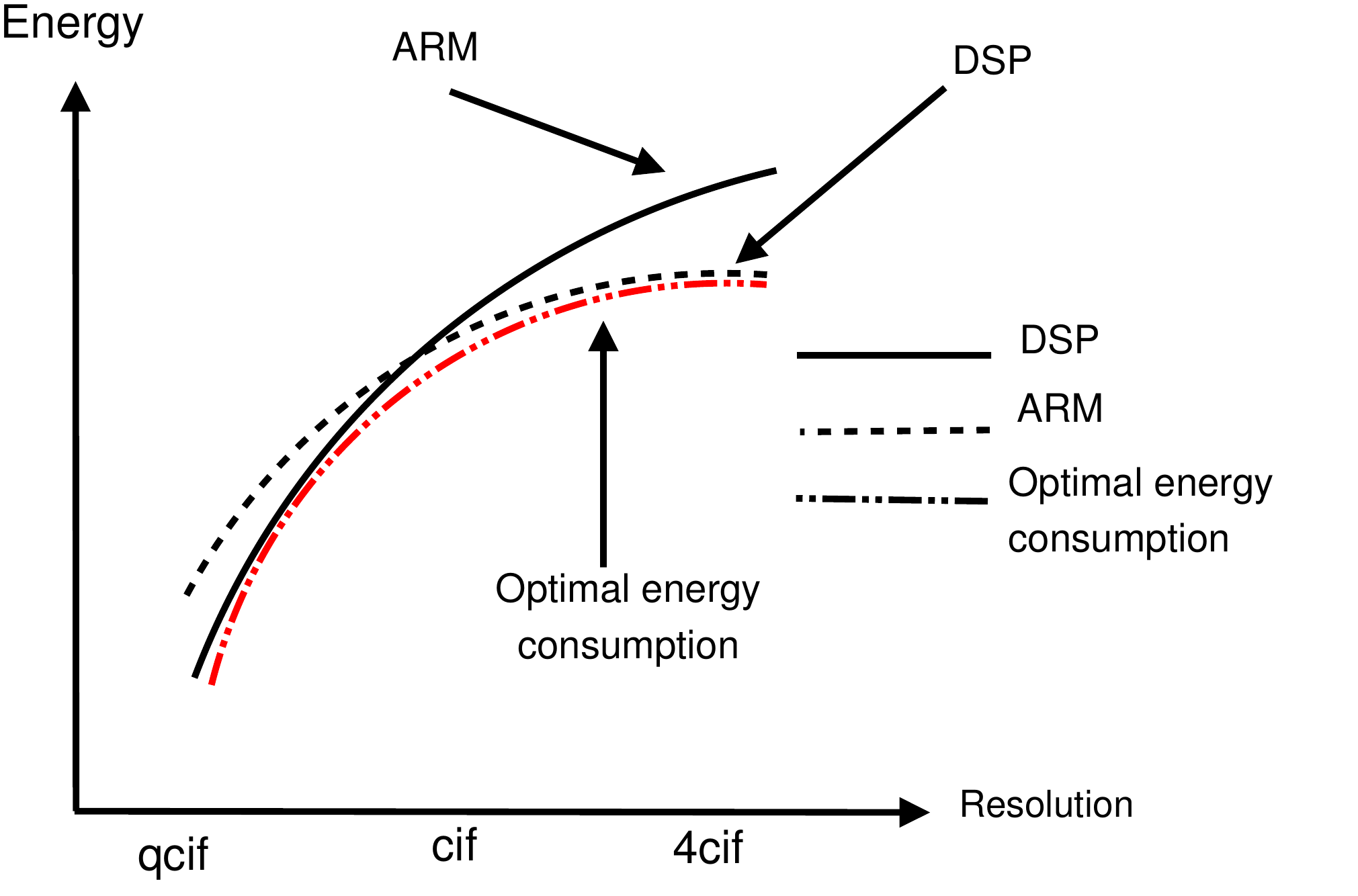}
 \caption{Optimal energy consumption according to resolution}
 \label{optim}
\end{figure}

Based on the observations discussed in section \ref{backgroud}, we propose  to implement a processor switching policy according to the video resolution. The implemented video player switches to ARM video decoding in case of low video resolutions (\textit{qcif}) and to DSP decoding  for higher resolution (\textit{cif}/\textit{4cif}). By doing so, the video decoder would select the best processing resources achieving the decoding task using the least energy as illustrated  in  Figure \ref{optim}.

We choose to implement such logic in case of dynamic adaptive streaming scenario because: 1) it is more used in real life as compared to scalable coding, 2) As discussed in section \ref{scope}, the different video qualities are independent from each other (unlike scalable coding) which allows to one video chunk to be decoded in a GPP and the next one on the DSP without worrying about any dependency issue. 

In what remains, we discuss the implementation details of the proposed dynamic processor switching technique in case of adaptive video streaming regardless of any existing video streaming technology and standard \cite{Stockhammer:2011:DAS:1943552.1943572,HLS,MSS}. This technique is implemented on an embedded Linux operating systems using the \textit{GStreamer} multimedia framework.

\section{DyPS Design and Implementation }
\label{implementation}
\subsection{Hardware Setup}

DyPS (Dynamic Processor Switching) was implemented on the OMAP3530 EVM board containing the low-power  OMAP3530 \textit{SoC} consisting of a Cortex A8 ARM processor and TMS320C64X DSP. The  power consumptions of the DSP and the ARM processors are measured using the Open-PEOPLE framework \cite{sennopenpeople}, a multi-user and multi-target power and energy optimization platform and estimator. It includes the NI-PXI-4472 digitizer allowing up to a 100 KHz sampling rate.


\subsection{Software Setup}

On this hardware platform, the Linux operating system version 2.6.32 was used. The H264/AVC video decoding was achieved using GStreamer \cite{gstreamer-omap35x}, a multimedia development framework. The ARM decoding, was performed using \textit{ffdec\_h264}, a plug-in based on the widely used  \textit{ffmpeg/libavcodec} library compiled with the support of NEON SIMD instructions set. For DSP decoding, we used \textit{TIViddec2}, a proprietary  H.264/AVC baseline profile plug-in provided by \textit{Texas Instrument}.  


	
\begin{figure}[!t]
\centering
\includegraphics[width=3.3in]{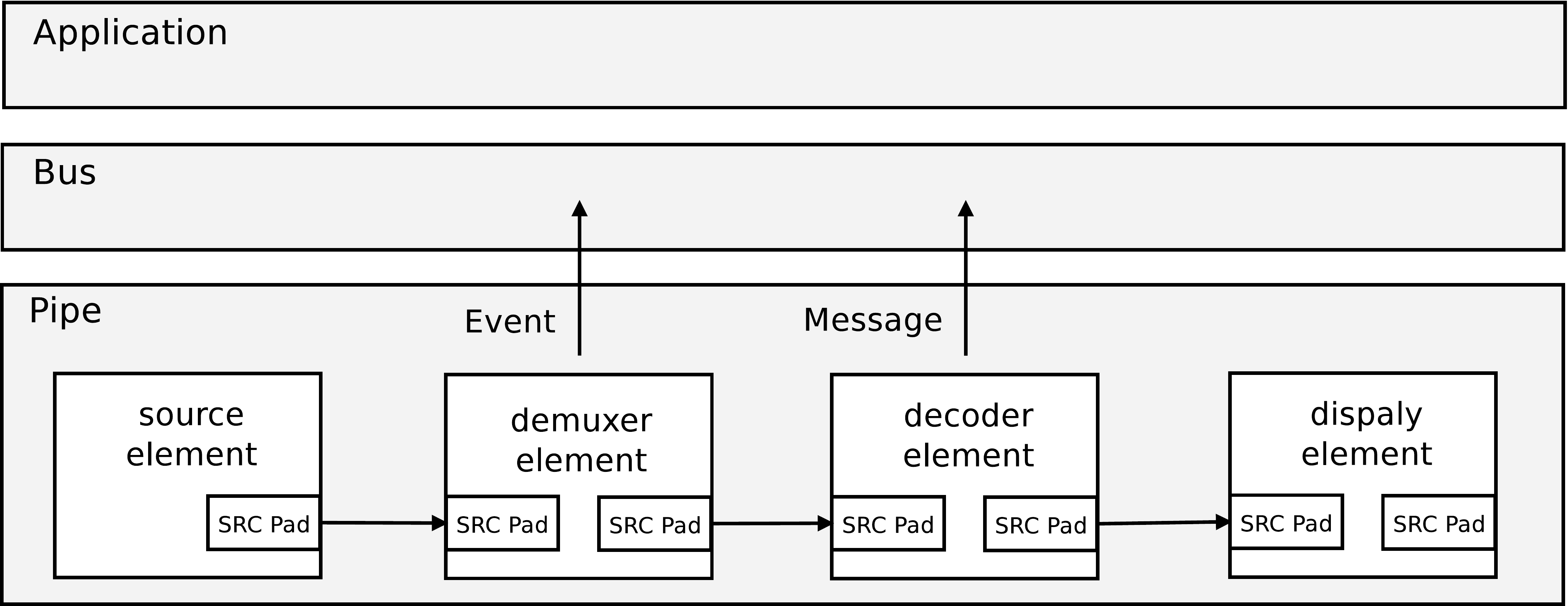}
 \caption{GStreamer framework}
 \label{gstreamer}
\end{figure}

GStreamer is a framework for creating streaming multimedia applications. It has a modular design based on plugins that  provide  various codecs and other functionalities. In \textit{GStreamer}, an \textit{Element} is the most important object. It has one specific function, which can be, for example, the reading of data from a file, decoding of this data or witting it to a display device. Many elements can be linked together to form a \textit{Pipe} and let data flow through this chain. As illustrated in Figure \ref{gstreamer}, In a typical decoding chain, the {Elements} are connected thanks to \textit{Pads} which are used to negotiate links and data flow between them.  Data flows out of one element through one or more source \textit{Pads}, and elements accept incoming data through one or more sink \textit{Pads}. The types of these data are described as a \textit{GstCaps} (Capabilities).

In addition, GStreamer provides powerful communication and synchronization mechanisms. Thus, different \textit{Elements} can exchange various types of messages through a \textit{Bus} and rise \textit{Events} which can be handled synchronously by a dedicated handler. The above-described GStreamer features make it suitable to implement our proposed solution since we have to deal with dynamic events related to video quality adaptation decoding using two types of codecs targeting a ARM and DSP processor.

\subsection{DyPS design}

In the proposed solution, we reproduce a typical adaptive streaming scenario where a video content is coded in different qualities  and divided into small chunks. Each chunk is coded using a video compression standard (in our case, H.264/AVC)  and contained in an MP4 file format. Thus, a complete decoding pipe  consists of the following elements: 

\begin{itemize}
\item \textit{filesrc} : for reading the video file. 
\item \textit{qtdemux} : for extracting the video content (H.264/AVC coded data) from the MP4 files. 
\item \textit{ffdec\_h264} or \textit{TIViddec2} : for decoding the coded H.264/AVC data using the ARM processor or the DSP. 
\item \textit{xvimagesink} : for displaying the video content. 

\end{itemize}

\begin{figure}[!h]
\centering
\includegraphics[width=3.4in]{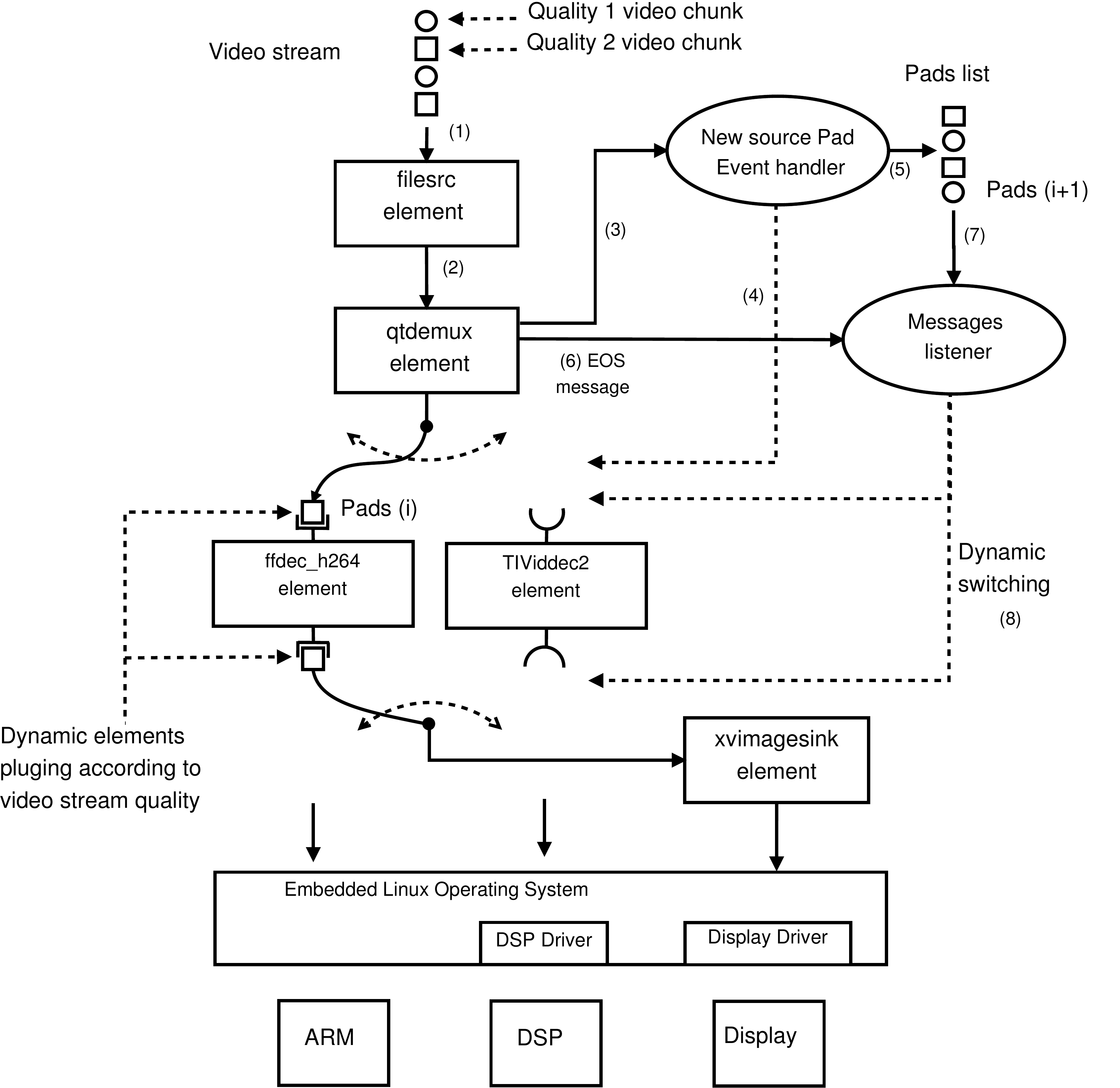}
 \caption{DyPS: Dynamic processor switching solution design using GStreamer}
 \label{dynamic-pluging}
\end{figure}

\noindent In DyPS, we suppose that different chunks are contained in one MP4 file. As illustrated in Fig. \ref{dynamic-pluging} : a video file is read using \textit{filesrc} element (1) then its video content is extracted (2) using \textit{qtdemux} demuxer. This element rises a "new pad" event (3) when it detects a new video chunk. According to the Pad type (in our case, the video resolution), a dedicated event handler plugs dynamically (4) the demuxer to the  \textit{ffdec\_h264} or \textit{TIViddec2} decoder element. The next detected pads are queued in a list (5). When a video chunk is totally played, an "End Of Stream" (EOS) message is sent via the communication bus (6). Each time an EOS is sent,  a message listener treats it by retrieving a pad from the list (7) and  plugs it to a decoder element (ARM or DSP decoder) according to the video quality (8). The processor switching  is achieved at this step. The selected decoder is then  connected to the \textit{xvimagesink} display element.
All these functionalities are controlled from the application using an API provided by the \textit{GStreamer} framework. 

\section{Case study}
\label{results}
\begin{figure}[!h]
\centering
\includegraphics[width=2in]{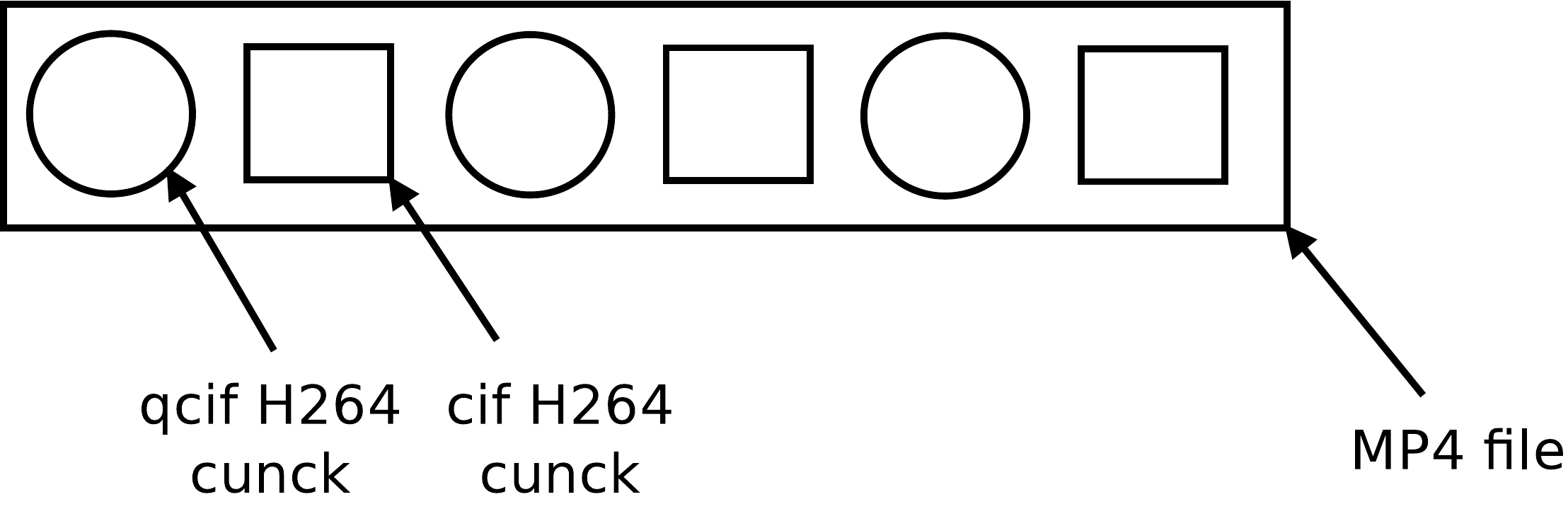}
 \caption{Video chunks in MP4 file}
 \label{chunks}
\end{figure}

As discussed in section \ref{adaptation-logic}, we have configured DyPS to play \textit{qcif} videos on ARM processor and higher resolution on DSP processor. 10 seconds H.264/AVC Harbor video chunks was grouped in a MP4 file as illustrated in Figure \ref{chunks}.   

\begin{figure}[!h]
\centering
\includegraphics[width=2.4in]{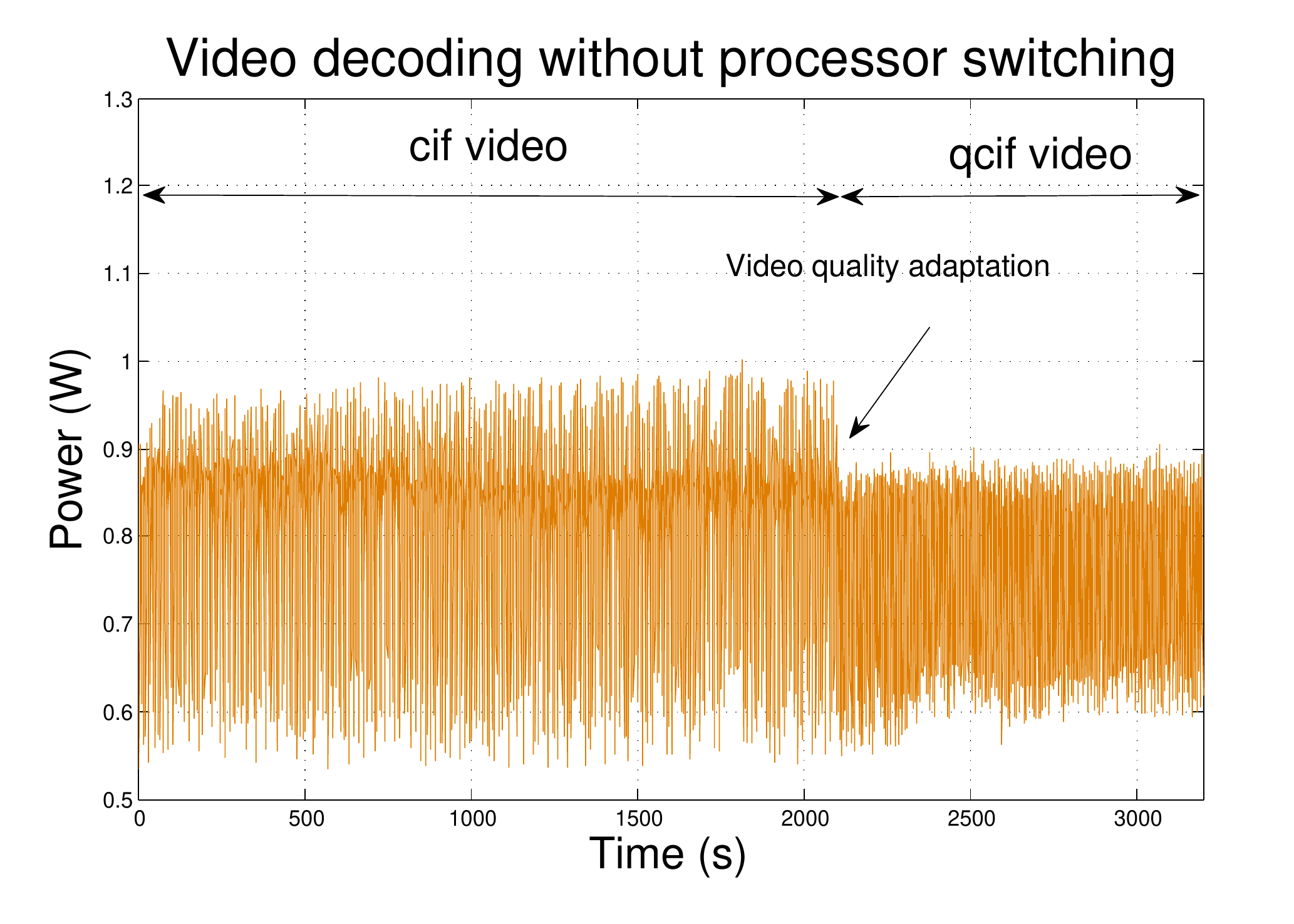}
\includegraphics[width=2.4in]{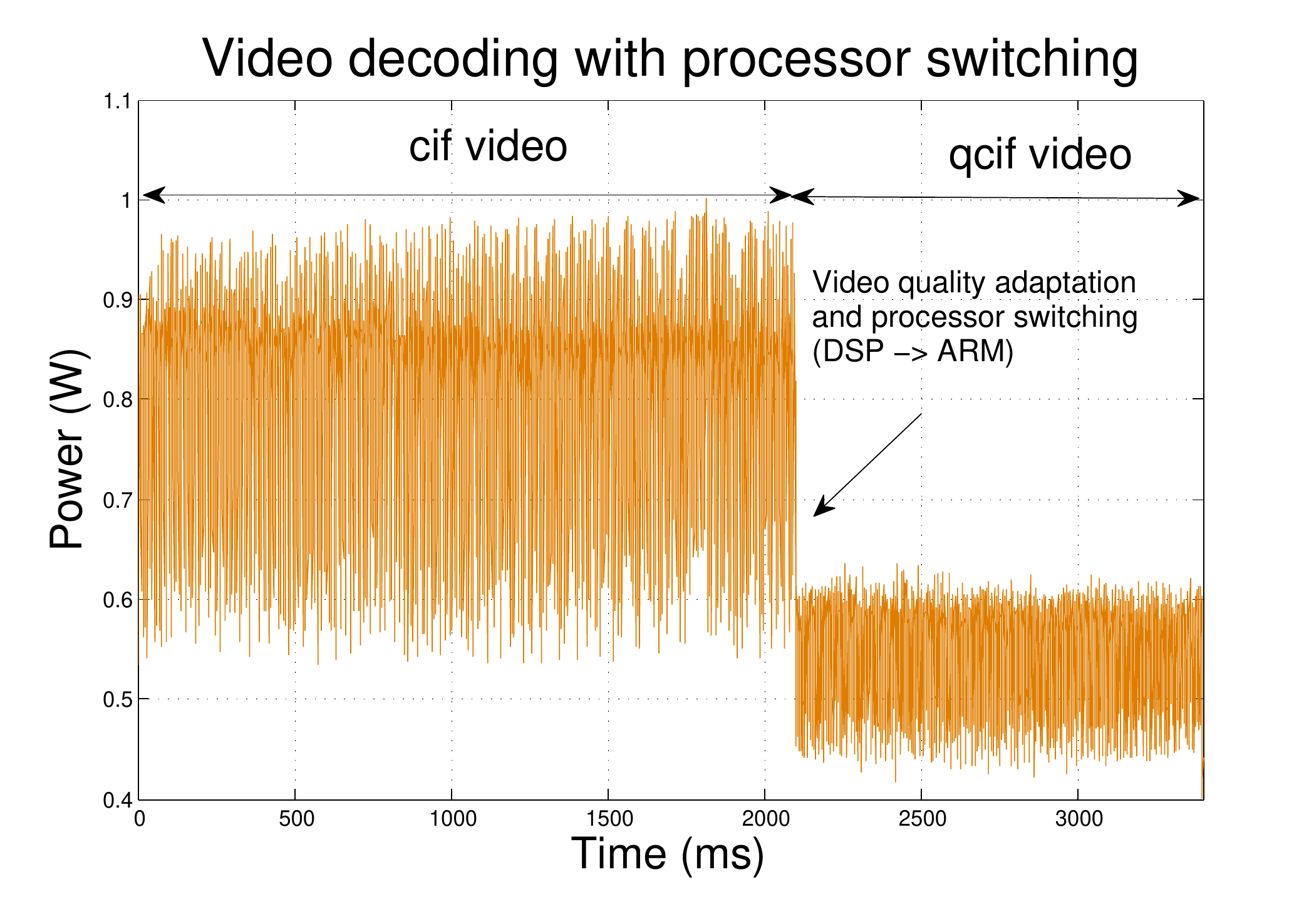}
 \caption{Dynamic processor switching impact on video decoding power consumption}
 \label{adaptation}
\end{figure}

The player (with DyPS plugged) achieved a transparent processor switching according to the decoded video resolution. Figure \ref{adaptation} shows the power consumption plots resulting from decoding consecutive (\textit{cif}, 4Mb/s)  and (\textit{qcif}, 512 Kb/s) chunks when disabling (Fig. \ref{adaptation}-a) and enabling (Fig. \ref{adaptation}-b) the processor switching. One can observe that switching to ARM decoding in case of \textit{qcif} resolution allows to reduce to power consumption. A 30\% energy saving is achieved in this example as compared to the DSP decoding when using DyPS without impact on the performances.

\section{Related Works}
\label{related}

Video decoding performance and energy consumption optimization issue has been addressed by many works. In \cite{1317052}, the performance and energy consideration when using pipelines and parallelism in CMOS circuits is studied and it was shown why these architectures increase the energy efficiency. In \cite{H264-ASIC}, the particular case of H.264 video decoding is analyzed and an energy-aware  architecture design methodology is proposed for energy efficient H.264/AVC decoding. At this level, the impact on the energy consumption of application and operating system layers are not considered. 
   
At a higher level, In \cite{1218201}, H.264/AVC decoding performance is characterized on different GPP processor architectures  at CPU cycle level. This approach is used  in \cite{macomplexity2011} for energy characterization and modeling of the different H.264/AVC decoder modules. The results were used to develop an energy-aware video decoding strategy for ARM processor supporting DVFS feature. The result of this study are generalized in  \cite{limodeling2012} for considering the variation in video bit-rate. These studies was focusing only on GPP processor. 

Many works studied the performance and energy consumption of DSP video decoding. In \cite{5657208,1598326,DSP-overhead-article}, performance consideration of DSP decoding are analyzed especially regarding cache coherency maintenance and DMA transfers. In \cite{Julien}, energy characterization of DSP processing is addressed in terms of memory access and DMA transfers. In \cite{juarez_distortion-energy_2012}, DSP video decoding energy consumption is analyzed in terms of different video coding qualities.  Many of these studies highlight the performance and the energy efficiency of the DSP video decoding.

In this work a combined GPP/DSP decoding technique in a context of adaptive video decoding was proposed width the objective to save energy. As far as we know, no study  proposed before such an approach. 

\section{Conclusion}
\label{conclusion}

In this paper, we described DyPS: a new technique for energy-aware dynamic processor switching based on energy consumption properties of GPP and DSP when decoding video. We have shown the benefit of using such a technique for energy saving in a context of dynamic video streaming. The feasibility of this technique was demonstrated by implementing it on the flexible and powerful GStreamer multimedia frameworks on embedded Linux platform. The processor switching criteria was based on the video resolution. This can be generalized to the video bit-rate. DyPS was validated on one hardware platform and a more thorough validation is to be performed on other platforms by characterizing the performance and energy consumption of GPPs and DSPs in order to confirm the  optimal energy consumption configuration (see Figure \ref{optim}).

A more elaborate policy can be envisaged for driving the processor switching technique. For example, the overall system load is an important criteria to be considered.  In fact, even ARM decoding is more energy efficient for some video qualities, the use of the DSP offloads the ARM processor and lets the operating system or other application  execute tasks without impacting the video playback. Thus, we can suggest to schedule a systematic DSP video decoding starting from a given system load threshold.

This work is a proof of concept and as a future works, we plan to implement this technique in a real adaptive streaming client and to extend the switching criteria to the bit-rate and the system load. 

\bibliographystyle{abbrv}

\bibliography{./bibliography}

\end{document}